%% file: main.tex
\documentclass[letterpaper,twocolumn,10pt]{article}
\usepackage{usenix}
\usepackage{tikz}
\usepackage{amsmath}
\usepackage{amsfonts}
\usepackage{enumitem}
\usepackage{algorithm}
\usepackage{algpseudocode}
\usepackage{tabularx}
\usepackage{booktabs}
\usepackage{caption}
\usepackage{graphicx}
\usepackage{textcomp}
\usepackage{multirow}
\usepackage{xspace}
\usepackage{filecontents}
\usepackage[english]{babel}
\usepackage{blindtext}
\usepackage{tikz}
\definecolor{comment}{rgb}{0.54,0.1,0.066}
\usepackage{todonotes}
\input{include}
\hypersetup{
  pdftitle={Slant: Shaping Traffic for Resource-Efficient Microservices},
  pdfauthor={Congwei Zhang; Chieh-Jan Mike Liang; Mengbai Xiao; Yifei Zou; Shulun Wu; Xiangyu Gao; Xiuzhen Cheng; Dongxiao Yu},
  pdfsubject={Resource-efficient traffic routing for microservice systems},
  pdfkeywords={Slant, microservices, routing, resource efficiency, autoscaling}
}

\newcommand{\paperauthor}[3]{%
  \begin{tabular}[t]{@{}c@{}}%
    {\normalsize\rmfamily #1\textsuperscript{#2}}\\[-1pt]
    {\scriptsize\ttfamily\href{mailto:#3}{#3}}%
  \end{tabular}%
}

\begin{document}

\date{}


\title{Slant: Shaping Traffic for Resource-Efficient Microservices}

\author{
  \setlength{\tabcolsep}{5pt}
  \begin{tabular}{@{}cccc@{}}
  \paperauthor{Congwei Zhang}{1}{xczhw@mail.sdu.edu.cn} &
  \paperauthor{Chieh-Jan Mike Liang}{2}{liang.mike@microsoft.com} &
  \paperauthor{Mengbai Xiao}{1}{xiaomb@sdu.edu.cn} &
  \paperauthor{Yifei Zou}{1}{yfzou@sdu.edu.cn} \\[1.5ex]
  \paperauthor{Shulun Wu}{1}{202335232@mail.sdu.edu.cn} &
  \paperauthor{Xiangyu Gao}{3}{gao-xy24@mails.tsinghua.edu.cn} &
  \paperauthor{Xiuzhen Cheng}{1}{xzcheng@sdu.edu.cn} &
  \paperauthor{Dongxiao Yu}{1}{dxyu@sdu.edu.cn}
  \end{tabular}\\[1.3ex]
  {\small\rmfamily
    \textsuperscript{1}Shandong University \quad
    \textsuperscript{2}Microsoft Research \quad
    \textsuperscript{3}Tsinghua University
  }
}
\maketitle

\input{0-Abstract}

\input{1-Introduction}
\input{2-Background}
\input{3-System}
\input{4-Implementation}
\input{5-Evaluation}

\input{6-RelatedWork}
\input{7-conclusion}

\clearpage
\bibliographystyle{unsrt} 
\bibliography{main}
\clearpage
\appendix

\end{document}

%% file: include.tex
\usepackage{amsmath}        
\usepackage{amsfonts}       
\usepackage{amsthm}         
\usepackage{mathtools}      

\usepackage{algorithm}      

\usepackage{graphicx}       
\usepackage{subcaption}     
\usepackage{float}          
\usepackage{booktabs}       
\usepackage{multirow}       
\usepackage{multicol}       
\usepackage{array}          
\usepackage{tabularx}       
\usepackage{makecell}       

\usepackage{xcolor}         
\usepackage{tikz}           
\usetikzlibrary{shapes, arrows, positioning, calc, fit, backgrounds}

\usepackage{listings}       
\usepackage{xurl}
\usepackage{microtype}      
\usepackage{enumitem}       
\usepackage{setspace}       
\usepackage{balance}        

\usepackage{siunitx}        
\usepackage{textcomp}       
\usepackage{gensymb}        

\theoremstyle{plain}
\newtheorem{theorem}{Theorem}[section]

\theoremstyle{definition}

\theoremstyle{remark}

\newtheorem{observationenv}[theorem]{Observation}
\newcommand{\observation}[1]{\begin{observationenv}#1\end{observationenv}}
\newcommand{\REQUIRE}{\Require}
\newcommand{\ENSURE}{\Ensure}
\newcommand{\STATE}{\State}
\newcommand{\FOR}{\For}
\newcommand{\ENDFOR}{\EndFor}

\newcommand{\RETURN}{\Return}

\newcommand{\COMMENT}{\textbf{comment}}

\newcommand{\Slant}{Slant}

%% file: 0-Abstract.tex
\begin{abstract}
Microservice architectures rely on load balancers to distribute requests across service replicas, yet existing designs largely optimize for balanced routing without considering how routing shapes autoscaling signals and resource efficiency. We present Slant, a resource-aware load balancing framework that intentionally creates safe traffic imbalance to improve resource utilization. Slant profiles service-specific critical points, computes conservative multi-dimensional capacity, derives imbalance-aware routing weights that maintain hot and cold pods, and falls back to safer behavior under anomalies. By exposing clearer utilization signals, Slant works with existing vertical pod autoscalers (VPAs) without modification. Experiments on three microservice benchmarks show that Slant reduces allocated CPU cores by up to 26\% while lowering P99 latency by up to 18\% compared to Round-Robin. Under dynamic workloads, Slant triggers VPA scale-up 45\% faster and completes scale-in 40\% faster. Under anomalies, it recovers 40\% faster from pod failures and limits latency amplification to 1.8$\times$ versus 2.6$\times$ during downstream slowdowns.
\end{abstract}

%% file: 1-Introduction.tex
\section{Introduction}
\label{sec:intro}
Microservice architecture has become the dominant paradigm for cloud-native applications~\cite{meta, alibaba, chi2026efficient}. In such systems, a single user request often traverses dozens of services before completion, supported by modern deployment and observability infrastructures such as Docker, gRPC, Istio, Prometheus, and Jaeger~\cite{docker, grpc, rest, istio, prometheus, grafana, jaeger, zhang2024distributed}. The load balancer serves as a critical component in this path, determining which backend instance handles each incoming request. The effectiveness of load balancing strategies directly impacts system latency, throughput, and resource utilization.

Existing load balancing and overload control strategies~\cite{lb-survey, lb1, lb2, lb3, lb4, lb5, lb6, lb9, lb10, sherry2025congestion, DBLP:conf/nsdi/KrishnaswamySBR22, addanki2022powertcp, taodistributed, zou2024distributed}, whether consistent-hashing-based designs~\cite{Maglev, ring-hashing, wu2025adaptive, zhang2026crisp}, production reverse proxies~\cite{nginx, envoy, traefik, yarp, ipvs, zhang2026unified}, or more sophisticated application-aware variants~\cite{Prequal, C3, zhang2026balancing}, still largely optimize for balanced or responsiveness-oriented dispatch. While this seems reasonable, these approaches share a fundamental blind spot: they focus on how traffic should be distributed at the moment of routing, without considering how routing decisions shape the resource signals observed by downstream managers. Load balancers and resource management systems operate largely independently, lacking coordination.

However, there exists an inherent coupling between these two systems that cannot be ignored. Auto-scalers, from threshold-based designs to learning-based systems, make decisions based on resource utilization metrics observed from each Pod~\cite{HPA, aws, EC2, autothrottle, Sinan, wang2026distributed}, and these metrics are directly shaped by the traffic distribution determined by the load balancer. In other words, the load balancing strategy directly influences the input signals that auto-scalers rely upon. This is not an optional association but an architectural inevitability: traffic distribution determines the load on each Pod, which in turn determines the metrics observed by the scaler, ultimately affecting scaling decisions. The problem is that existing load balancing strategies are designed without any consideration of this downstream impact.

\begin{figure}[t]
    \centering
    \includegraphics[width=\linewidth]{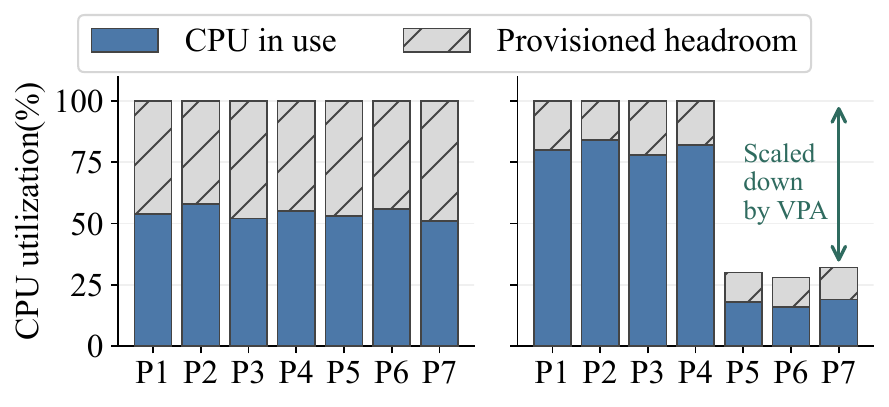}
    \caption{Balanced traffic obscures scaling signals and fragments reserved capacity, while intentional imbalance preserves similar workload with substantially lower total allocation.}
    \label{fig:motivation}
\end{figure}

This disconnect leads to tangible efficiency losses. When a load balancer evenly distributes traffic across all instances, each Pod operates at an ambiguous utilization level, typically hovering around 50--60\%. For auto-scalers, this represents the most challenging input scenario: scaling up appears unnecessary while scaling down seems risky. The result is conservative inaction. Figure~\ref{fig:motivation} illustrates this effect using seven Pods (P1--P7): in both bar charts, the orange portion represents used CPU and the gray portion represents allocated but unused CPU. Under balanced traffic distribution, all Pods maintain similar moderate utilization and retain similar CPU allocation, providing no clear signal for scaling decisions. Meanwhile, each instance must reserve headroom to handle potential traffic bursts, but this headroom is scattered across the entire fleet and cannot be consolidated, leading to systematic resource fragmentation. In contrast, under skewed traffic, a few Pods run hot while several cold Pods remain lightly loaded and can be assigned much smaller CPU limits, allowing the system to sustain a similar amount of useful work with substantially less total allocated CPU.

The deeper issue is that load balancing and auto-scaling pursue objectives that are not naturally aligned. Load balancing seeks instantaneous balance across instances, while vertical auto-scaling seeks efficient resource utilization over time~\cite{autothrottle, Sinan}. These goals can conflict. Consider a scenario where the VPA determines that a Pod's CPU limit can be safely reduced to save resources. The ideal condition for this operation is a Pod with consistently low load that clearly indicates over-provisioning. However, a balance-oriented load balancing strategy continuously assigns traffic to this Pod, keeping it in an ambiguous state where scaling down never becomes safe to execute.

Our key insight is that balanced traffic distribution does not equate to efficient resource utilization. In certain conditions, intentionally creating imbalance yields better outcomes. By directing traffic such that some instances operate at higher utilization while others remain lightly loaded, we achieve several benefits: heavily loaded instances emit clear signals for resource demand, and lightly loaded instances can have their resources scaled down safely by VPA. The critical question becomes: how much imbalance is safe? The answer depends on service-specific performance characteristics and resource sensitivity, which are known to vary substantially across systems and operating conditions~\cite{Prequal, autothrottle, zeng2025learning}. Different services exhibit vastly different tolerance to resource pressure, and even within a single service, different resource dimensions such as CPU, memory, and I/O have different safety boundaries.

Through our investigation, we make the following observations that guide our design. First, balanced traffic distribution obscures scaling signals, making it difficult for both threshold-based and learning-based auto-scalers to reach confident decisions. Second, performance critical points vary significantly across services due to differences in their computational characteristics and resource dependencies. Third, different resource dimensions exhibit different tolerance characteristics, with CPU allowing graceful degradation while memory exhibits cliff behavior upon exceeding limits. We elaborate on these observations with detailed measurements in Section~\ref{sec:background}.

Based on these observations, we propose \Slant, a resource-aware load balancing framework that leverages intentional traffic imbalance to maximize resource efficiency while satisfying SLO requirements and coordinating with existing auto-scalers. At a high level, Slant combines a \emph{Service Profiler} to learn service-specific performance boundaries, a \emph{Multi-dimensional Constraint Solver} to estimate safe remaining capacity across resource dimensions, an \emph{Imbalance-aware Routing Optimizer} to periodically compute routing weights that concentrate traffic on selected hot pods while preserving cold pods, and a \emph{Fallback Controller} to detect anomalies and revert to safer behavior when needed. These control-plane decisions are continuously fed by monitoring signals and executed by a lightweight data plane, enabling Slant to improve scaling signals without modifying the underlying auto-scaler. Our main contributions are as follows:

\begin{itemize}[leftmargin=*]
    \item We identify and analyze the overlooked coupling between load balancing and auto-scaling, demonstrating that existing balance-oriented strategies inadvertently create difficulties for resource management.
    \item We propose the concept of intentional imbalance and develop Slant, a complete framework that profiles service critical points, enforces multi-dimensional safety constraints, computes imbalance-aware routing weights, and provides fallback control under anomalies.
    \item We show how Slant works with existing vertical auto-scalers by creating clearer scale-out and scale-in signals, enabling faster and more decisive resource adjustment without changing the auto-scaler itself. On Social Network, Slant triggers the first VPA scale-up 45\% faster and completes scale-in 40\% faster than conventional balanced routing.
    \item We evaluate Slant on multiple microservice benchmarks and anomaly scenarios. Slant reduces allocated CPU cores by up to 26\% while also lowering P99 latency by up to 18\% and reducing SLO violation minutes by 60\%. Under failures and downstream slowdowns, it recovers 40\% faster from pod failure and limits latency amplification to 1.8$\times$ versus 2.6$\times$ for Round-Robin.
\end{itemize}

%% file: 2-Background.tex
\section{Background and Motivation}
\label{sec:background}

Our goal is to improve resource efficiency for microservice applications while satisfying end-to-end latency SLOs. The motivating measurements in this section are conducted on Social-Network from DeathStarBench~\cite{deathstarbench_repo}, deployed on a smaller Kubernetes cluster with 160 CPU cores across five nodes. We use Locust to generate workloads following real-world diurnal patterns, with a 200ms P99 latency SLO. Our full end-to-end evaluation in Section~\ref{sec:evaluation} is performed on a larger 20-node cluster.

\subsection{Balanced Distribution Obscures Scaling Signals}
\label{sec:obs1}

Auto-scalers rely on resource utilization metrics to make scaling decisions. Learning-based Vertical Pod Autoscaler (VPA) approaches like Autothrottle~\cite{autothrottle} learn the relationship between CPU throttle ratios and latency to adjust resource limits. They assume that observed metrics accurately reflect true resource pressure.

However, balanced load distribution causes all replicas to operate at similar, moderate utilization levels, creating ambiguity for auto-scalers. Consider a service with four replicas each at 55\% CPU utilization. This presents a weak signal: the auto-scaler cannot easily distinguish genuine resource pressure from transient fluctuations, leading to conservative resource allocation.

\begin{figure}[t]
    \centering
    \includegraphics[width=0.9\linewidth]{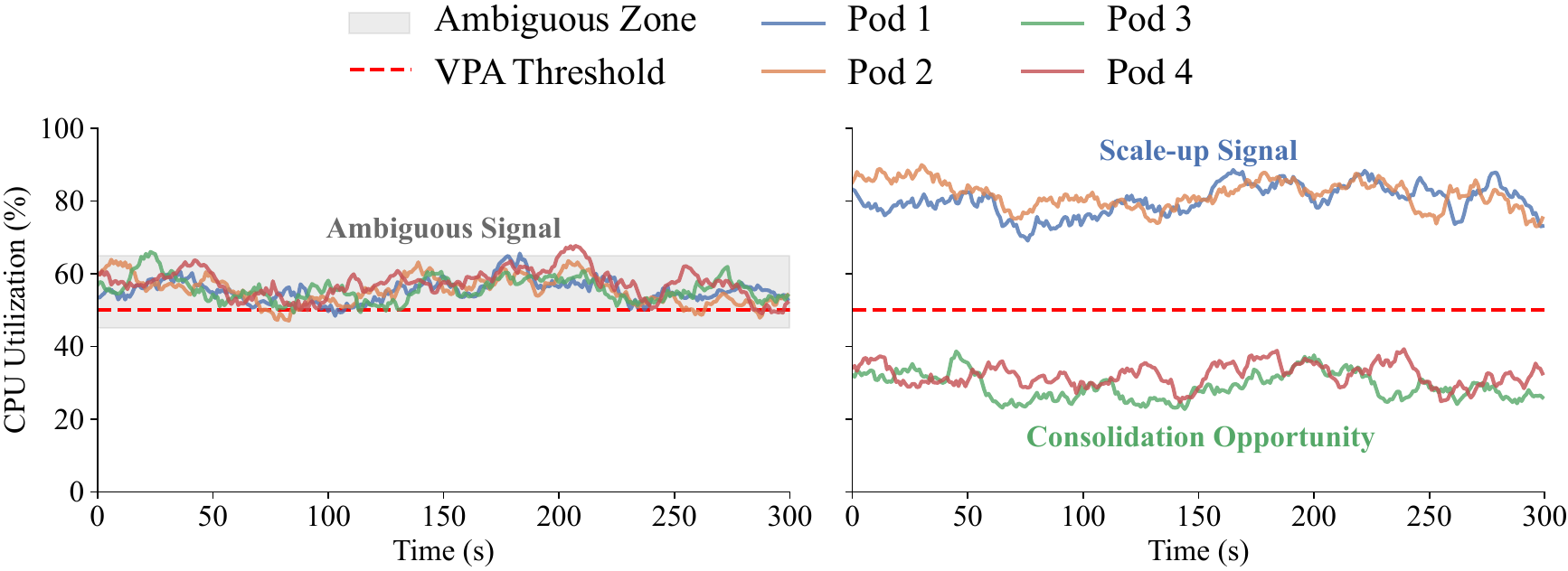}
    \caption{CPU utilization under balanced versus imbalanced distribution. Balanced distribution obscures scaling signals.}
    \label{fig:obs1}
\end{figure}

Figure~\ref{fig:obs1} illustrates this phenomenon for the \texttt{compose-post} service. Under Round-Robin distribution, all replicas hover around 50--60\% utilization, providing no clear scaling signal for VPA to reduce allocations. Under imbalanced distribution (70\% traffic to two replicas, 30\% to others), utilization divergence becomes apparent: heavily loaded replicas signal actual capacity needs, while lightly loaded ones clearly indicate safe scale-down opportunities for the VPA.

\vspace{0.5em}
\observation{Balanced traffic distribution causes similar utilization across replicas, preventing auto-scalers from identifying clear scaling opportunities.}






\subsection{Performance Critical Points Vary Across Services}
\label{sec:obs3}

Different microservices exhibit different performance degradation patterns under resource pressure. Some degrade gracefully until roughly 85\% utilization; others experience sharp latency increases at 60--70\% due to contention or queuing effects.

\begin{figure}[t]
    \centering
    \includegraphics[width=0.9\linewidth]{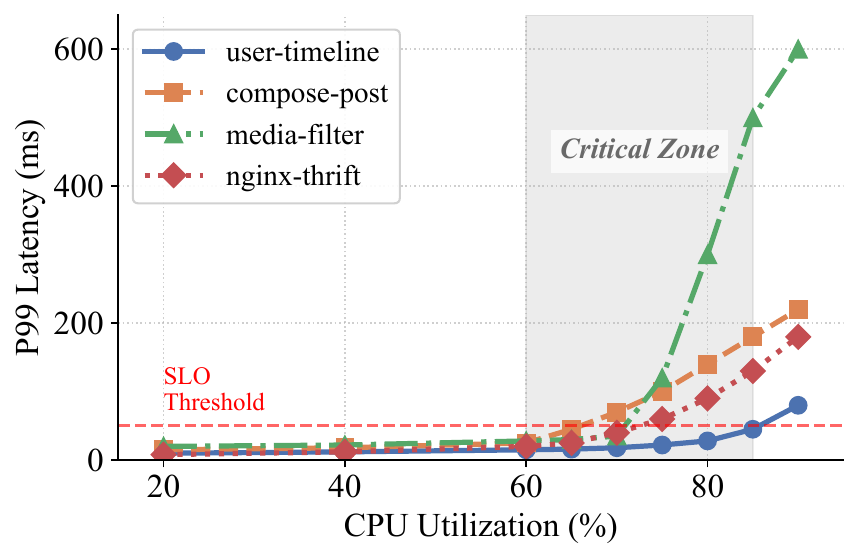}
    \caption{P99 latency versus CPU utilization for four services. Critical points range from 60\% to 85\%.}
    \label{fig:obs3}
\end{figure}

Figure~\ref{fig:obs3} profiles the latency-utilization relationship for four services. The \texttt{user-timeline} service shows gentle degradation, while \texttt{compose-post} has a sharp inflection at 65\% utilization. The \texttt{media-filter} ML service displays cliff behavior at 75\%. These differences arise from varying compute patterns, synchronization requirements, and external dependencies.

\vspace{0.5em}
\observation{Microservices exhibit different degradation patterns, with performance critical points varying from 60\% to 85\% utilization.}

\subsection{Resource Dimensions Differ in Tolerance}
\label{sec:obs4}

CPU and memory exhibit fundamentally different behavior under pressure. CPU causes gradual degradation, meaning the CFS scheduler throttles processes and increases latency predictably. Memory pressure remains stable until the limit is reached, then triggers GC pauses or OOM kills with catastrophic latency impact.

\begin{figure}[t]
    \centering
    \includegraphics[width=0.9\linewidth]{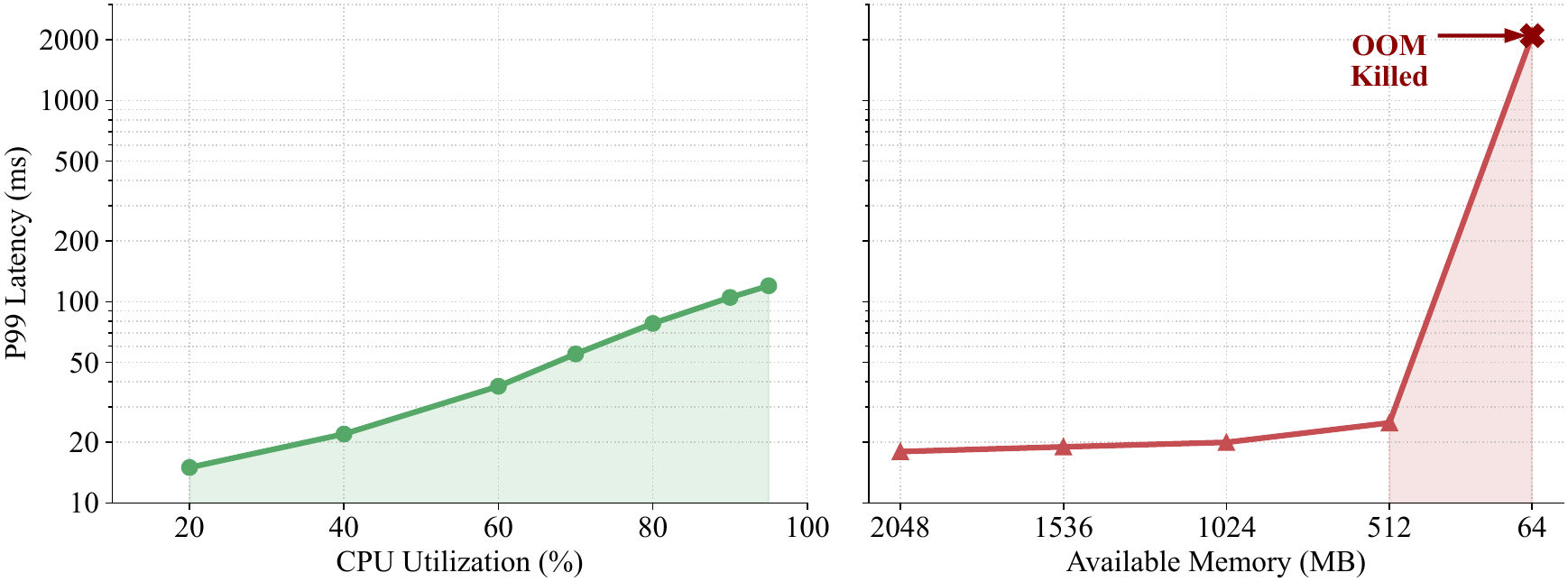}
    \caption{Latency under CPU (left) versus memory (right) pressure. CPU degrades gradually; memory exhibits cliff behavior.}
    \label{fig:obs4}
\end{figure}

Figure~\ref{fig:obs4} shows this for \texttt{post-storage-mongodb}. CPU pressure (left) increases latency smoothly from 15ms to 120ms. Memory pressure (right) maintains 20ms latency until dropping below 512MB, then spikes to over 2 seconds.

\vspace{0.5em}
\observation{CPU allows gradual degradation suitable for reactive scaling; memory exhibits cliff behavior requiring larger safety margins.}

\subsection{Design Implications}

Our observations reveal that balanced distribution obscures scaling signals, fragments headroom, and ignores heterogeneous service characteristics. This motivates intentional imbalance, which is carefully calibrated to per-service critical points and multi-dimensional constraints. Such an approach provides clearer signals, consolidates headroom, and maintains SLO compliance.

%% file: 3-System.tex
\section{System Design}
\label{sec:design}


\subsection{System Overview}
\label{subsec:overview}


\begin{figure}[t]
    \centering
    \includegraphics[width=\columnwidth]{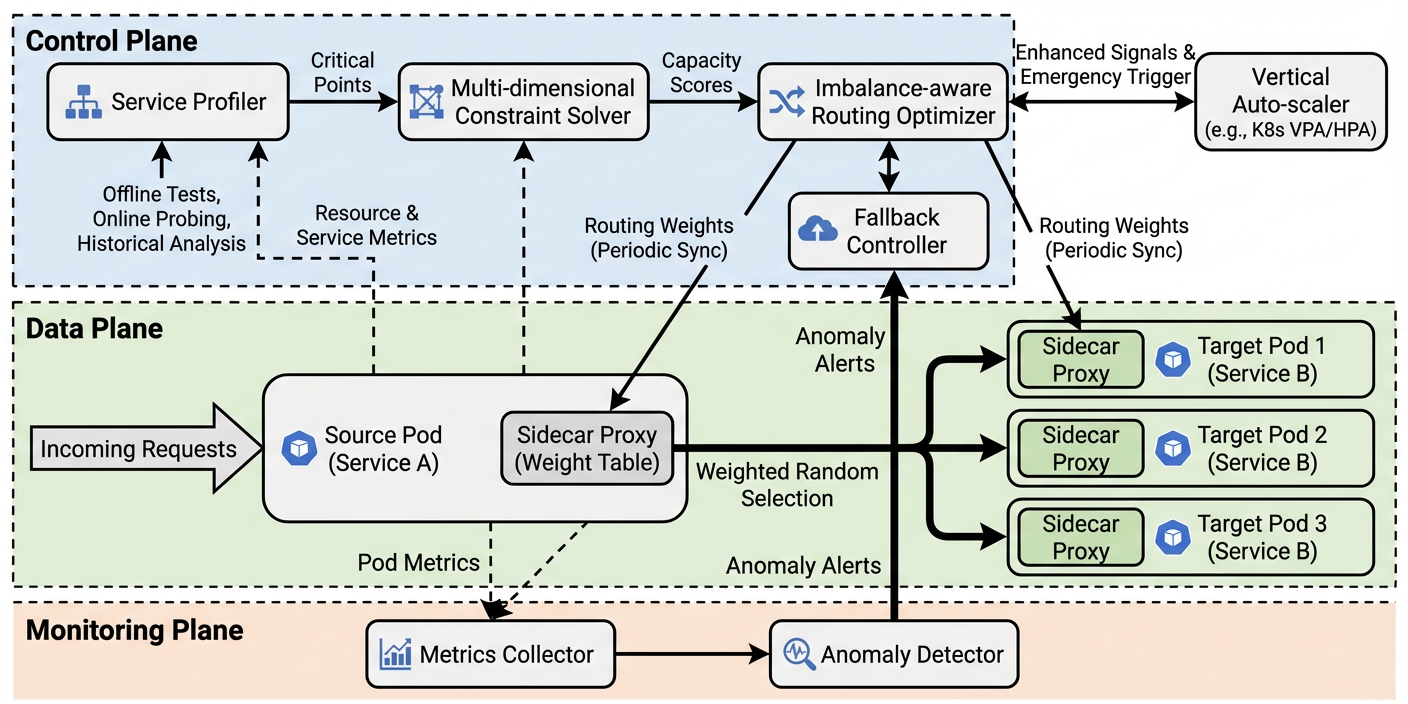} 
    \caption{Architecture overview of Slant. The system consists of three planes: the control plane computes routing decisions, the data plane executes request forwarding, and the monitoring plane collects metrics and detects anomalies.}
    \label{fig:architecture}
\end{figure}

Figure~\ref{fig:architecture} illustrates the overall architecture of Slant. The system is organized into three planes that work together to achieve imbalance-aware load balancing.

\paragraph{Control Plane.}
The control plane serves as the brain of Slant, responsible for computing and coordinating routing decisions. It consists of four core components. The \emph{Service Profiler} builds and maintains a multi-dimensional profile for each service, characterizing its critical points across different resource dimensions. The \emph{Multi-dimensional Constraint Solver} takes the profiled critical points and real-time resource metrics as input, computing the remaining capacity of each pod while ensuring safety across all resource dimensions. The \emph{Imbalance-aware Routing Optimizer} uses these capacity estimates to compute optimal routing weights that intentionally create load imbalance by concentrating traffic on some pods while keeping others lightly loaded. The \emph{Fallback Controller} monitors for anomalies and triggers corrective actions when the system deviates from expected behavior.

The control plane operates on a periodic basis (every 5 seconds by default) to update routing weights. This design choice balances responsiveness with stability: frequent updates enable quick adaptation to load changes, while avoiding the oscillation that per-request decisions might cause.

\paragraph{Data Plane.}
The data plane handles the actual request forwarding. We implement the data plane logic as a sidecar proxy co-located with each service instance, following the service-mesh deployment model popularized by systems such as Istio and Envoy~\cite{istio, envoy}. Upon receiving an incoming request, the sidecar consults a local weight table (periodically synchronized from the control plane) and selects a target pod using weighted random selection. The data plane is intentionally kept lightweight. It performs no complex computation, relying solely on table lookups and random number generation. This design ensures minimal latency overhead to request processing.

\paragraph{Monitoring Plane.}
The monitoring plane continuously collects resource metrics (CPU, memory, I/O, network) and service-level metrics (latency percentiles, error rates) from all pods. These metrics serve two purposes. First, they feed into the control plane to inform routing decisions. Second, they enable real-time anomaly detection. When the anomaly detector identifies signs of overload or degradation, it triggers the Fallback Controller to take corrective action.

\paragraph{Interface with Vertical Auto-scalers.}
A key design goal of Slant is to work seamlessly with existing auto-scaling infrastructure. Rather than replacing the auto-scaler, we enhance the signals it receives. By intentionally creating load imbalance, Slant causes some pods to exhibit high utilization (clearly signaling the need for scaling CPU limits up) while others remain near-idle (clearly eligible for scaling CPU limits down). Standard vertical auto-scalers and learning-based resource managers such as Autothrottle and Sinan can then make more decisive scaling decisions based on these clearer signals, without any modification to the auto-scaler itself~\cite{autothrottle, Sinan}.

\subsection{Service Profiler}
\label{subsec:profiler}

The Service Profiler addresses a fundamental challenge: determining how much load each service can safely handle. Different services have vastly different critical points, which denote the utilization level beyond which performance degrades unacceptably. A stateless computation service might tolerate 85\% CPU utilization without issues, while a service with complex lock contention might start suffering at 50\%. The Profiler's job is to learn these critical points for each service across multiple resource dimensions.

\subsubsection{Critical Point Modeling}

We model the critical point not as a single threshold but as a probability distribution. This reflects the reality that performance degradation is not a sharp cliff but a gradual slope, and that the exact point of degradation varies with workload characteristics, time of day, and other factors.

Formally, for service $s$ and resource dimension $d$ (e.g., CPU, memory), we model the critical point as a random variable $C_{s,d}$ following a distribution parameterized by $(\mu_{s,d}, \sigma_{s,d})$. The probability that utilization level $u$ is safe can be expressed as:

\begin{equation}
    P(\text{safe} \mid u, s, d) = P(C_{s,d} > u) = 1 - \Phi\left(\frac{u - \mu_{s,d}}{\sigma_{s,d}}\right),
\end{equation}

where $\Phi$ is the cumulative distribution function of the standard normal distribution. When making routing decisions, we operate at a conservative percentile (e.g., the 10th percentile of the distribution), ensuring high confidence that we remain within safe bounds.

\subsubsection{Hybrid Profiling Methodology}

The Profiler employs three complementary methods to estimate and update critical point distributions, as each method has distinct strengths and limitations.

\paragraph{Offline Stress Testing.}
Before a service enters production, we conduct controlled stress tests in a staging environment. We gradually increase the load while monitoring latency percentiles (P50, P99, P999) and error rates. The critical point is identified as the utilization level where latency exceeds acceptable thresholds (e.g., P99 latency increases by more than 50\% from baseline) or errors begin to appear. This method provides accurate initial estimates but cannot capture production-specific behaviors or long-term drift.

\paragraph{Online Probing.}
In production, we occasionally and carefully probe the boundaries of known-safe regions. When a pod's utilization approaches the estimated critical point with high headroom in other pods (ensuring we can absorb traffic if problems arise), we allow slightly higher utilization and observe the outcome. Successful probes (no degradation observed) cause us to slightly increase our estimate of $\mu_{s,d}$; degradation events cause us to decrease it. We implement this as a conservative bandit algorithm that prioritizes safety over exploration.

\paragraph{Historical Analysis.}
We continuously analyze historical data correlating resource utilization with service-level indicators (latency, errors). Using changepoint detection algorithms, we identify instances where performance degraded and correlate them with resource metrics at that time. This passive observation complements active probing by capturing degradation events that occur naturally, including rare edge cases that deliberate probing might miss.

\subsubsection{Handling Profile Drift}

Service characteristics change over time due to code updates, data growth, and shifting traffic patterns. The Profiler addresses this through continuous adaptation and explicit invalidation. The online probing and historical analysis methods continuously refine estimates, allowing gradual drift to be tracked automatically. For abrupt changes (e.g., a new deployment), we detect the event through integration with the CI/CD pipeline and temporarily widen the confidence interval (increase $\sigma_{s,d}$) until sufficient new observations accumulate.

\subsection{Multi-dimensional Constraint Solver}
\label{subsec:constraint}

The Multi-dimensional Constraint Solver addresses a critical insight: different resource dimensions have fundamentally different characteristics and failure modes. CPU overload causes slowdown but is recoverable; memory exhaustion causes crashes with no graceful degradation. A routing decision that looks safe from a CPU perspective might be dangerous from a memory perspective. The Solver's role is to compute a unified ``remaining capacity'' score for each pod that respects constraints across all dimensions.

\subsubsection{Per-dimension Capacity Modeling}

For each pod $p$ and resource dimension $d$, we compute the remaining capacity as the gap between current utilization and the conservative critical point:

\begin{equation}
    \text{Cap}_{p,d} = C^{\text{conservative}}_{s(p),d} - U_{p,d},
\end{equation}

where $s(p)$ denotes the service that pod $p$ belongs to, $C^{\text{conservative}}_{s,d}$ is the conservative (e.g., 10th percentile) critical point from the Profiler, and $U_{p,d}$ is the current utilization of pod $p$ on dimension $d$.

Different dimensions require different safety margins based on their failure characteristics. For CPU, we allow aggressive utilization up to the critical point (typically 75--85\%), since CPU overload manifests as increased latency but recovers quickly when load decreases. For memory, we maintain a substantial safety margin (typically 25--35\% below the limit), because memory pressure can trigger garbage collection storms, and exceeding limits causes immediate OOM kills with no opportunity for graceful handling. For I/O, safety margins depend on the storage medium, such that SSDs tolerate higher utilization than HDDs. We also account for I/O queue depth, as high queue depths cause latency spikes even at moderate throughput. For network, we consider both bandwidth utilization and connection counts, since high connection counts can exhaust file descriptors or cause connection tracking table overflow.

\subsubsection{Unified Capacity Score}

The key insight of the Solver is that a pod's true remaining capacity is determined by its most constrained dimension. We compute a unified capacity score as:

\begin{equation}
    \text{Cap}_p = \min_{d \in \mathcal{D}} \left( w_d \cdot \text{Cap}_{p,d} \right),
\end{equation}

where $\mathcal{D}$ is the set of resource dimensions and $w_d$ is a normalization weight that accounts for different scales across dimensions. This min-based aggregation ensures that routing decisions respect all constraints simultaneously, treating a pod with 50\% CPU headroom but only 5\% memory headroom as nearly full.

\subsubsection{Request Impact Estimation}

A remaining challenge is that the resource impact of an incoming request is unknown before it executes. We address this through statistical modeling of request costs. For each service, we maintain a distribution of per-request resource consumption across dimensions, estimated from historical observations. In the current system, this model is aggregated into the periodic capacity estimates used by the routing optimizer; when constructing these estimates, we consider the expected resource impact plus a safety margin:

\begin{equation}
    \text{Impact}_d = \mathbb{E}[\text{cost}_d] + \alpha \cdot \text{Std}[\text{cost}_d],
\end{equation}

where $\alpha$ controls the conservativeness (typically 1--2 standard deviations). These impact estimates are folded into the per-pod capacity scores used for periodic weight computation, ensuring that routing weights are derived from conservative multi-dimensional feasibility estimates rather than raw utilization alone.

For services with heterogeneous request types, Slant can be extended to be request-type-aware. If request metadata (e.g., API endpoint or query parameters) is available, the control plane can maintain separate cost distributions per request class and derive class-specific capacity estimates and weight tables. We leave this request-type-aware execution path as an extension rather than a required component of the current implementation.

\subsection{Imbalance-aware Routing Optimizer}
\label{subsec:routing}

The Routing Optimizer is the core component that realizes our key insight: intentionally creating load imbalance to improve resource efficiency and scaling clarity. Unlike traditional load balancers that aim to equalize utilization across pods, our optimizer deliberately concentrates load on some pods while keeping others lightly loaded.

\subsubsection{Optimization Formulation}

We formulate routing weight computation as an optimization problem. Let $N$ be the number of pods for a service, $\lambda$ be the total incoming request rate, and $w_i$ be the routing weight for pod $i$ (where $\sum_i w_i = 1$). Let $\text{Cap}_i$ denote the unified capacity score from the Constraint Solver.

The objective function balances utilization maximization with maintaining cold pods:

\begin{equation}
    \max_{\mathbf{w}} \sum_{i=1}^{N} \left( U_i(\mathbf{w}) + \beta \cdot \mathbf{1}[U_i(\mathbf{w}) < \tau_{\text{cold}}] \right)
\end{equation}

subject to:
\begin{align}
    U_i(\mathbf{w}) &\leq C^{\text{conservative}}_i, \quad \forall i \\
    \sum_{i=1}^{N} w_i &= 1, \quad w_i \geq 0,
\end{align}

where $U_i(\mathbf{w}) = w_i \cdot \lambda \cdot \text{cost}_{\text{avg}}$ is the expected utilization of pod $i$ given routing weights $\mathbf{w}$, $\tau_{\text{cold}}$ is the threshold for considering a pod ``cold'' (e.g., 10\% utilization), and $\beta$ is a reward for maintaining cold pods to facilitate scale-in.

\subsubsection{Practical Algorithm}

Solving the full optimization problem at each decision epoch can be computationally expensive. We employ a practical heuristic algorithm that achieves near-optimal results efficiently.

\begin{algorithm}[t]
\caption{Imbalance-aware Weight Computation}
\label{alg:routing}
\begin{algorithmic}[1]
\REQUIRE Pods $\{p_1, \ldots, p_N\}$, capacity scores $\{\text{Cap}_1, \ldots, \text{Cap}_N\}$, target cold pod ratio $\gamma$
\ENSURE Routing weights $\{w_1, \ldots, w_N\}$

\STATE Sort pods by current utilization: $U_{(1)} \leq U_{(2)} \leq \cdots \leq U_{(N)}$
\STATE $N_{\text{cold}} \leftarrow \lceil \gamma \cdot N \rceil$ \COMMENT{Number of cold pods to maintain}
\STATE $\mathcal{C} \leftarrow \{p_{(1)}, \ldots, p_{(N_{\text{cold}})}\}$ \COMMENT{Cold pod set}
\STATE $\mathcal{H} \leftarrow \{p_{(N_{\text{cold}}+1)}, \ldots, p_{(N)}\}$ \COMMENT{Hot pod candidates}

\FOR{$p_i \in \mathcal{C}$}
    \STATE $w_i \leftarrow \epsilon$ \COMMENT{Minimal weight for cold pods}
\ENDFOR

\STATE $W_{\text{remaining}} \leftarrow 1 - |\mathcal{C}| \cdot \epsilon$

\FOR{$p_i \in \mathcal{H}$}
    \STATE $w_i \leftarrow W_{\text{remaining}} \cdot \frac{\text{Cap}_i}{\sum_{p_j \in \mathcal{H}} \text{Cap}_j}$
\ENDFOR

\STATE \textbf{Verify:} Ensure $U_i(w_i) \leq C^{\text{conservative}}_i$ for all $i$
\STATE \textbf{Adjust:} If any constraint violated, redistribute excess weight

\RETURN $\{w_1, \ldots, w_N\}$
\end{algorithmic}
\end{algorithm}

Algorithm~\ref{alg:routing} describes our weight computation procedure. The key idea is to partition pods into two groups: cold pods that should receive minimal traffic, and hot pods that should absorb the remaining load. Among hot pods, traffic is distributed proportionally to their remaining capacity, ensuring that pods with more headroom receive more traffic.

The algorithm runs periodically (every 5 seconds) rather than per-request, striking a balance between adaptiveness and stability. Per-request decisions would enable finer-grained control but risk oscillation and add latency overhead.

\subsubsection{Coordination with Auto-scaler}

The routing strategy creates utilization patterns that naturally enhance auto-scaler effectiveness. When incoming traffic increases, the hot pods approach their capacity limits. Because we concentrate load on fewer pods (rather than spreading thinly across all), these pods generate clear resource-pressure signals earlier and more decisively. The VPA therefore observes stronger evidence that CPU limits should be increased.

When traffic decreases, our algorithm prioritizes draining specific pods to near-zero utilization while maintaining reasonable load on others. The VPA then observes clear scale-down candidates whose CPU limits can be reduced safely, rather than a fleet of moderately utilized pods that all look ambiguous.

\subsection{Fallback Controller}
\label{subsec:fallback}

The Fallback Controller provides safety guarantees when things go wrong. Despite careful profiling and conservative decision-making, prediction errors are inevitable. Traffic patterns may shift unexpectedly, service behavior may change due to bugs, or cascading failures may occur in downstream dependencies. The Controller's role is to detect problems quickly and take corrective action before they escalate.

\subsubsection{Anomaly Detection}

The Controller continuously monitors service-level indicators to detect signs of distress. We track three primary signals. For latency deviation, we maintain an exponentially weighted moving average (EWMA) of P99 latency for each pod; when current P99 exceeds the EWMA by more than a threshold (e.g., 50\%), we flag the pod as potentially overloaded. We use relative deviation rather than absolute thresholds to handle services with naturally different latency profiles. For error rate spikes, a sudden increase in error rate (e.g., HTTP 5xx responses, timeout errors) indicates severe problems; we monitor the error rate over a sliding window and trigger alerts when it exceeds baseline by a significant margin. For resource exhaustion warnings, even before latency degrades, we may observe leading indicators of impending problems: rapid memory growth approaching limits, CPU throttling events, or I/O queue depth spikes. These early warnings enable proactive intervention.

\subsubsection{Graduated Response}

When anomalies are detected, the Controller responds with graduated actions based on severity. This design prevents overreaction to transient issues while ensuring serious problems are addressed promptly.

At Level 1 (Targeted Traffic Shift), for mild anomalies such as latency slightly elevated on one pod, we reduce the routing weight for the affected pod while redistributing its share to healthy pods. This surgical intervention often resolves the issue without broader disruption. If metrics recover within a timeout period (e.g., 30 seconds), we gradually restore normal weights.

At Level 2 (Fallback to Balanced), if Level 1 intervention fails to resolve the issue, or if multiple pods show signs of distress simultaneously, we temporarily abandon the imbalanced strategy and revert to round-robin distribution. This provides breathing room by spreading load evenly, reducing the risk of any single pod being overwhelmed. We maintain this fallback state until metrics stabilize, then cautiously re-engage imbalanced routing.

At Level 3 (Emergency Escalation), for severe situations such as high error rates across multiple pods or rapid resource exhaustion, we take aggressive action. First, we enable rate limiting at the service entry point to cap incoming traffic at a sustainable level. Second, we send an explicit signal to the auto-scaler requesting emergency scale-out, bypassing normal threshold-based triggers. Third, we alert operators for manual investigation.

\subsubsection{Recovery and Learning}

After a fallback event, the system does not simply resume normal operation; instead, it learns from the incident. The Profiler receives feedback about the conditions that led to the anomaly, allowing it to adjust critical point estimates. If a pod showed distress at 75\% CPU utilization when our estimate predicted safety up to 80\%, we revise $\mu_{s,\text{CPU}}$ downward accordingly.

We also implement a ``cooling off'' period after incidents. The system operates more conservatively (lower target utilization, wider safety margins) for a configurable duration, ensuring stability before returning to efficiency-optimized operation.

%% file: 4-Implementation.tex
\section{Implementation}
\label{sec:implementation}

We implement Slant as a standalone system that integrates with Kubernetes-based microservice deployments. The current prototype consists of approximately 8,500 lines of Go code for the control plane and 1,200 lines of Lua code for the data plane extensions. The control plane runs as a Kubernetes service that periodically computes routing weights from monitored resource signals, while the data plane is implemented through Envoy sidecars that execute lightweight weighted forwarding decisions. Detailed descriptions of the experimental testbed, workloads, and metrics are deferred to Section~\ref{subsec:methodology}.

%% file: 5-Evaluation.tex
\section{Evaluation}
\label{sec:evaluation}
\subsection{Methodology}
\label{subsec:methodology}

\paragraph{Benchmark Applications.}
We evaluate Slant using three diverse microservice applications that represent different complexity levels and architectural patterns commonly found in production environments.

\emph{Online Boutique} is a lightweight e-commerce application from Google comprising 10 microservices with simple gRPC communications~\cite{Online-boutique}. Most request flows involve only 2 to 3 services with minimal fan-out, making it ideal for measuring baseline performance and algorithmic overhead in low-complexity environments.

\emph{Social Network}, adapted from DeathStarBench~\cite{deathstarbench_repo}, represents moderate complexity with 28 microservices including both stateless and stateful components. It features two ML-based services that introduce computational intensity and hardware acceleration dependencies. With typical request paths spanning 4 to 6 services, this workload effectively tests adaptive algorithms under conditions of service heterogeneity and mixed resource demands.

\emph{Train Ticket}, developed by Fudan University~\cite{Train-ticket}, simulates a comprehensive ticket booking system with 68 microservices. This compute-intensive application includes complex dependencies, request chains exceeding 10 services, and mixed RPC communications. Its large scale and varying latency profiles make it particularly valuable for stress-testing load balancing policies under high concurrency.

All three applications run on Kubernetes within an Istio service mesh, ensuring a consistent environment that allows for direct comparison of load balancing behaviors across different workloads.

\paragraph{Trace-Driven Workload.}
Beyond the benchmark applications, we also evaluate Slant under a 24-hour production trace derived from an anonymized large-scale web service. This trace-driven experiment is used to study long-running VPA behavior under diurnal dynamics and bursty demand, complementing the controlled benchmark-based comparisons above.

\paragraph{Comparison Baselines.}
We compare Slant against four baseline load balancing strategies:

\begin{itemize}[leftmargin=*]
    \item \emph{Round-Robin (RR):} Distributes requests equally across all available pods in a circular fashion. This is the default strategy in many production deployments due to its simplicity.

    \item \emph{Least-Connections (LC):} Routes each request to the pod with the fewest active connections, providing basic load awareness without resource monitoring.

    \item \emph{Power of Two Choices (P2C):} A classic randomized load balancing strategy that randomly selects two candidate pods and routes the request to the one with fewer active connections, avoiding the herd effect of LC.

    \item \emph{Prequal}~\cite{Prequal}: A state-of-the-art adaptive load balancer that selects servers based on estimated latency and active requests-in-flight (RIF). It actively probes server load and leverages the power-of-d-choices paradigm, extended with asynchronous reusable probes, representing the best existing approach for microservice environments.
\end{itemize}

All baselines use the same Envoy-based data plane infrastructure to ensure fair comparison, differing only in their weight computation logic.

\paragraph{Experiment Setup.}
We conduct experiments on a Kubernetes cluster with 20 worker nodes, each equipped with 32 CPU cores (Intel Xeon Gold 6326 @ 2.90GHz), 128GB RAM, and 1TB NVMe SSD, connected via 25Gbps Ethernet. We use Kubernetes 1.28 and deploy Autothrottle~\cite{autothrottle, cloudlab} as our VPA, dynamically adjusting CPU limits to meet P99 latency SLOs.

For load generation, we use a distributed setup with dedicated client machines running customized wrk2~\cite{wrk2} instances. Unless otherwise specified, we generate load following an open-loop model to accurately measure latency under varying conditions. Each experiment runs for at least 10 minutes after a 2-minute warmup period, and we report the average of three independent runs.

We define an SLO violation as any request with end-to-end latency exceeding the 99th percentile target: 100ms for Online Boutique, 200ms for Social Network, and 500ms for Train Ticket. These thresholds are based on typical production requirements for similar applications.

\subsection{End-to-End Performance}
\label{subsec:e2e}

We first evaluate the overall performance of Slant compared to baselines under steady-state conditions. For each benchmark application, we generate constant load at 60\% of the maximum sustainable throughput and measure latency distribution, throughput, and resource utilization.


\begin{table}[t]
    \centering
    \caption{End-to-end performance summary at 60\% load. Slant achieves the best resource efficiency (lowest CPU cores allocated) while maintaining competitive latency and zero SLO violations.}
    \label{tab:e2e-summary}
    \small
    \begin{tabular}{llccc}
        \toprule
        \textbf{App} & \textbf{System} & \textbf{P99 (ms)} & \textbf{CPU Cores} & \textbf{SLO Viol.} \\
        \midrule
        \multirow{5}{*}{\rotatebox{90}{Boutique}}
        & Round-Robin & 52 & 48.5 & 0.02\% \\
        & Least-Conn & 49 & 48.0 & 0.01\% \\
        & P2C & 48 & 47.5 & 0.01\% \\
        & Prequal & 46 & 47.0 & 0.01\% \\
        & Slant & \textbf{45} & \textbf{43.5} & \textbf{0.00\%} \\
        \midrule
        \multirow{5}{*}{\rotatebox{90}{Social}}
        & Round-Robin & 156 & 115.0 & 0.15\% \\
        & Least-Conn & 148 & 113.5 & 0.12\% \\
        & P2C & 142 & 112.0 & 0.08\% \\
        & Prequal & 139 & 110.5 & 0.05\% \\
        & Slant & \textbf{128} & \textbf{81.5} & \textbf{0.02\%} \\
        \midrule
        \multirow{5}{*}{\rotatebox{90}{Train}}
        & Round-Robin & 412 & 285.5 & 0.28\% \\
        & Least-Conn & 395 & 282.0 & 0.22\% \\
        & P2C & 378 & 278.5 & 0.18\% \\
        & Prequal & 362 & 275.0 & 0.12\% \\
        & Slant & \textbf{338} & \textbf{210.5} & \textbf{0.05\%} \\
        \bottomrule
    \end{tabular}
\end{table}

Table~\ref{tab:e2e-summary} summarizes the end-to-end performance across all benchmark applications. On Online Boutique, all systems perform similarly due to the application's simplicity, but Slant still delivers the lowest P99 latency (45ms) and the fewest allocated CPU cores (43.5). On Social Network, Slant reduces P99 latency by 12\% relative to Round-Robin and 8\% relative to Prequal, while cutting allocated CPU cores from 115.0 to 81.5. On Train Ticket, the most complex benchmark, Slant lowers P99 latency from 412ms to 338ms and reduces allocated CPU cores from 285.5 to 210.5. Overall, Slant consistently requires the fewest allocated CPU cores, saving up to 26\% compared to baselines, while also delivering the lowest P99 latency and fewest SLO violations. These results show that intentional load imbalance does not sacrifice service quality; instead, it improves both resource efficiency and performance by operating pods near their efficient region and consolidating scattered headroom.

\begin{figure*}[t]
    \centering
    \includegraphics[width=\textwidth]{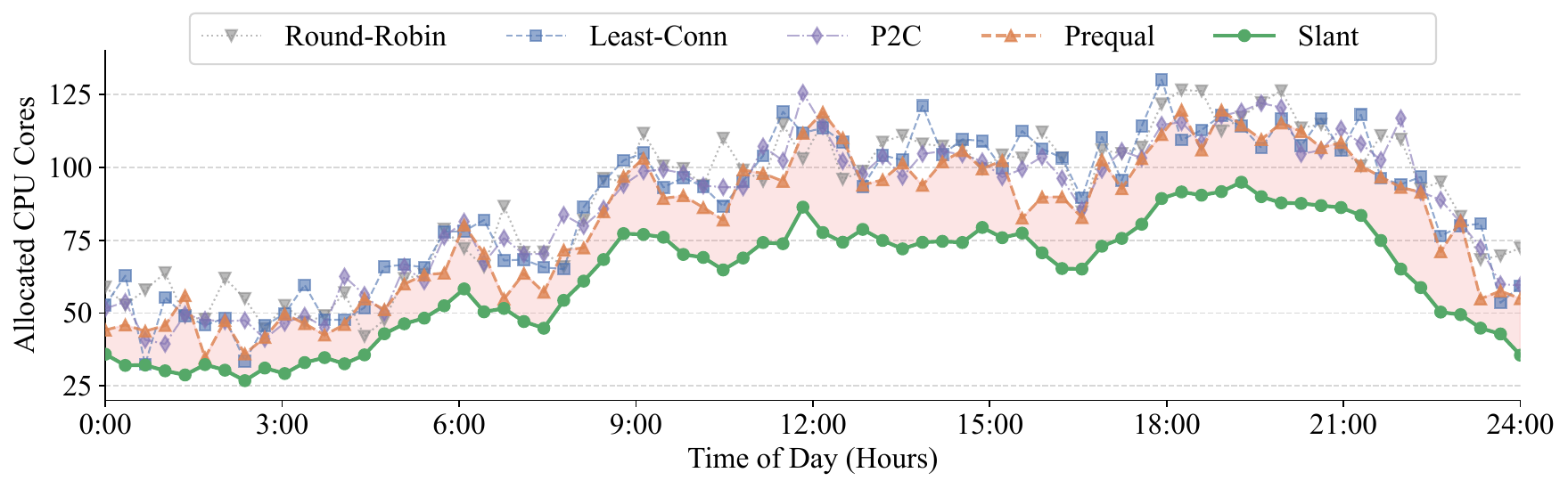} 
    \caption{Resource allocation under a 24-hour production trace. Slant enables the VPA to track load tightly, significantly reducing over-provisioned cores while minimizing SLO violations.}
    \label{fig:trace-autoscaling}
\end{figure*}

\subsection{Auto-scaling Effectiveness}
\label{subsec:autoscaling}

A key hypothesis of Slant is that intentional load imbalance produces clearer signals for auto-scalers like VPA, enabling faster and more accurate resource adjustments. We evaluate this hypothesis using dynamic workloads that require the system to scale resource limits up and down.

\subsubsection{Response to Step-load Variations}

We first evaluate how quickly each system triggers resource adjustments under sudden load shifts. Starting from a steady state at 40\% capacity, we instantaneously increase load to 80\% capacity at $t=0$, hold it, and then drop it back to 40\% at $t=300$ seconds. This allows us to measure both scale-out speed (provisioning sufficient CPU cores to prevent SLO violations) and scale-in efficiency (reclaiming over-provisioned cores to save resources).


\begin{figure}[t]
    \centering
    \includegraphics[width=\columnwidth]{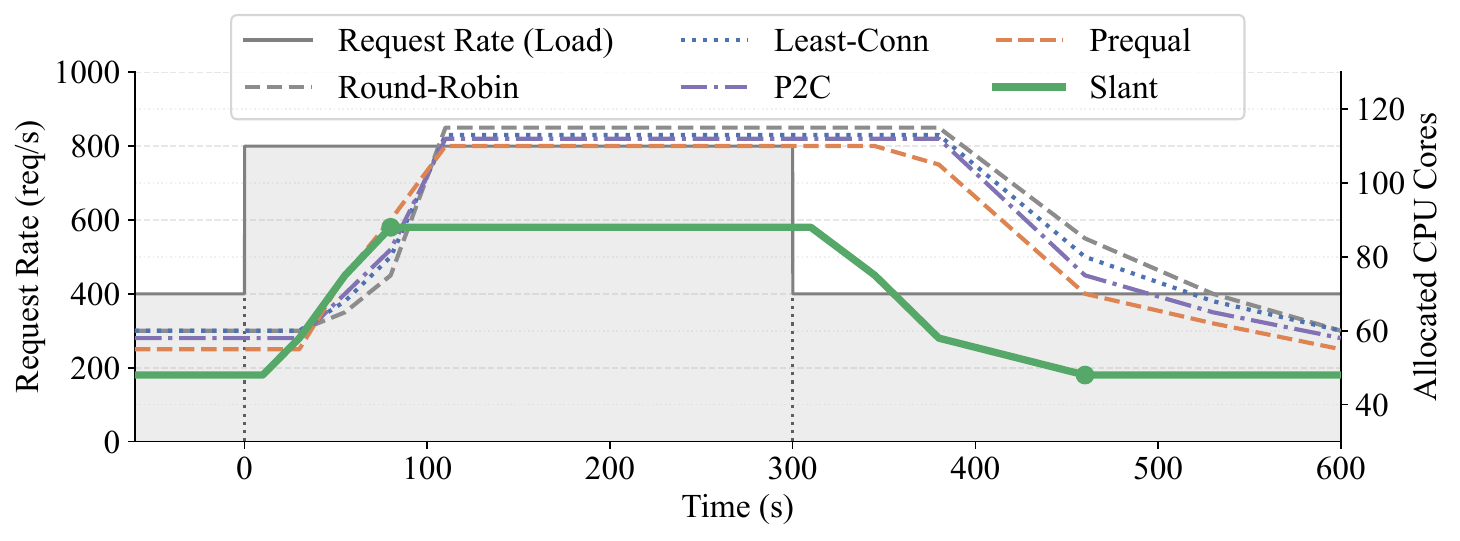} 
    \caption{VPA response to step-load variations. Slant accelerates both scale-out and scale-in by creating clear resource pressure and under-utilization signals.}
    \label{fig:stepload-timeline}
\end{figure}

Figure~\ref{fig:stepload-timeline} shows the resource adjustment timeline on Social Network. During the scale-out phase ($t=0$), Slant triggers the first VPA scale-up event within 18 seconds, while Round-Robin takes 33 seconds. This 45\% improvement occurs because Slant concentrates load on hot pods, causing them to exhibit high latency and CPU throttling much faster. Round-Robin, conversely, spreads load evenly, so all pods slowly drift toward saturation without clearly signaling urgent demand. Slant reaches the target CPU core allocation in 52 seconds, compared to 78 seconds for Round-Robin. During this scaling period, Slant experiences only 0.8\% SLO violations (vs. 2.3\% for Round-Robin) because cold pods provide immediate overflow capacity.

During the scale-in phase ($t=300$), traditional load balancers face a fundamental challenge: balanced distribution causes all pods to drop to moderate utilization, leaving the VPA uncertain about how much resource can be safely reclaimed. Slant addresses this by actively draining specific pods to near-zero utilization. These pods become unambiguous scale-in candidates, allowing the VPA to act decisively. Consequently, Slant completes scale-in 40\% faster than baselines, significantly reducing the time excess CPU cores consume cluster resources.

\subsubsection{Dynamic Workload Trace}

We then evaluate auto-scaling performance under a 24-hour diurnal trace from a production environment, applying continuous, dynamic load. Figure~\ref{fig:trace-autoscaling} shows the CPU cores allocated over time. Because Slant continuously creates clear resource pressure signals for VPA scale-up and clear under-utilization signals for VPA scale-in, its resource allocation curve tracks the actual load curve much more tightly than baselines.


Over the 24-hour period, Slant achieves a 22\% reduction in average CPU cores allocated compared to Prequal, and 28\% compared to Round-Robin. Despite using fewer resources, Slant experiences 60\% fewer SLO violation minutes than Round-Robin, demonstrating that tighter VPA tracking improves both efficiency and reliability.

\subsection{Resource Efficiency under Varying Load}
\label{subsec:efficiency}

We now examine how resource efficiency varies across different load levels. To make the tradeoff clearer, we separate resource efficiency into two views: the number of pods required to sustain each load level, and the resulting average CPU utilization.

\begin{figure}[t]
    \centering
    \begin{subfigure}[t]{\columnwidth}
        \centering
        \includegraphics[width=\linewidth]{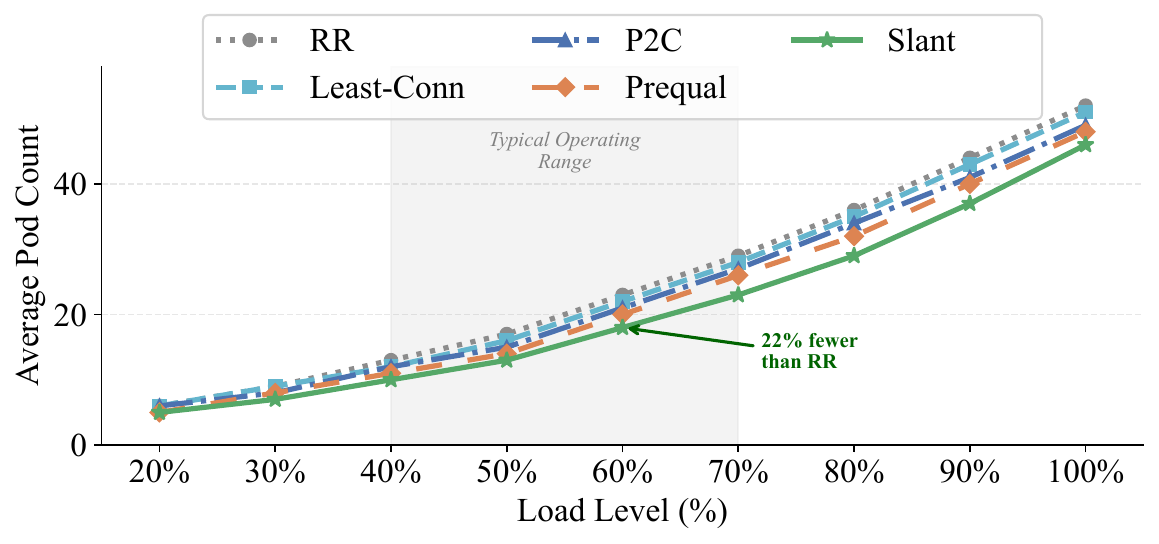}
        \caption{Average pod count.}
        \label{fig:load-pods}
    \end{subfigure}
    \hfill
    \begin{subfigure}[t]{\columnwidth}
        \centering
        \includegraphics[width=\linewidth]{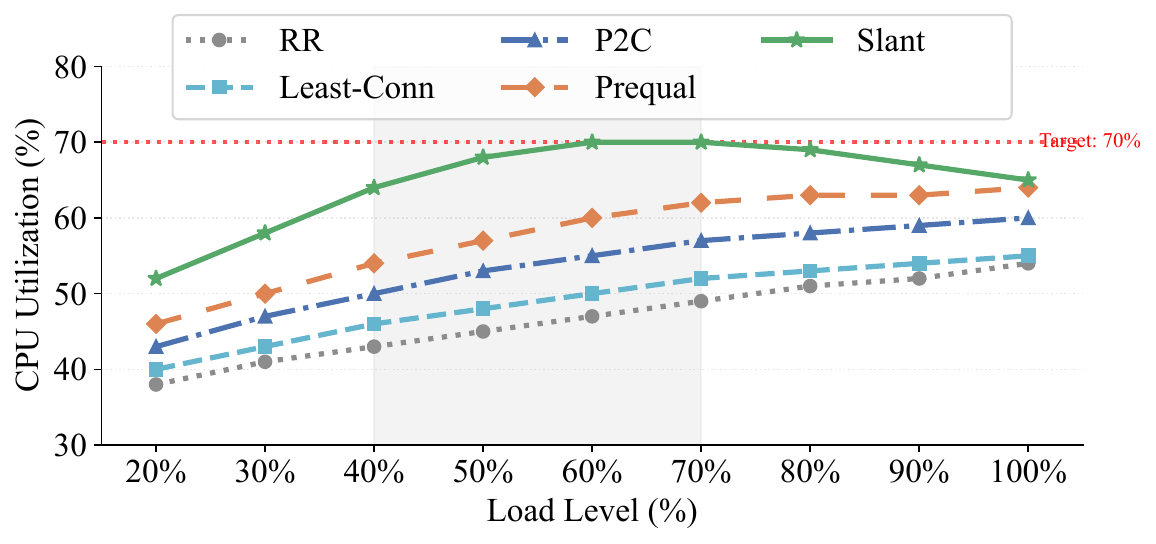}
        \caption{Average CPU utilization.}
        \label{fig:load-cpu}
    \end{subfigure}
    \caption{Resource efficiency across load levels on Social Network. Slant consistently requires fewer pods while maintaining higher CPU utilization. P2C improves over Least-Conn, but remains less efficient than Prequal and Slant.}
    \label{fig:load-efficiency}
\end{figure}

Figure~\ref{fig:load-efficiency} summarizes how different policies trade off provisioned capacity and achieved utilization on Social Network. The left subfigure reports the average number of pods needed at each load level, while the right subfigure reports the corresponding average CPU utilization.

First, Slant consistently requires the fewest pods across the entire load range. At 60\% load, Slant uses 18 pods, compared to 23 for Round-Robin, 22 for Least-Conn, 21 for P2C, and 20 for Prequal. This corresponds to a 22\% reduction relative to Round-Robin and still a 10\% reduction relative to the strongest baseline, Prequal.

Second, the CPU utilization plot shows the complementary effect of the same behavior. Baselines such as Round-Robin and Least-Conn spread requests more evenly, which leaves more partially utilized pods and keeps average utilization in the 40--56\% range. P2C and Prequal improve this trend, but Slant drives utilization substantially higher, reaching 58--70\% and staying close to the 70\% target over the typical operating region.

Third, the two subfigures together highlight the efficiency ordering among methods: Round-Robin is the least efficient, Least-Conn improves modestly, P2C provides a stronger baseline, Prequal performs better still, and Slant is best overall. The advantage is most visible in the moderate load regime (40--70\%), where production services spend much of their time. At very low load, minimum pod-count constraints limit the gap, while at very high load all methods begin to converge because sufficient total capacity must still be provisioned to avoid SLO violations.


\begin{figure}[t]
    \centering
    \includegraphics[width=\columnwidth]{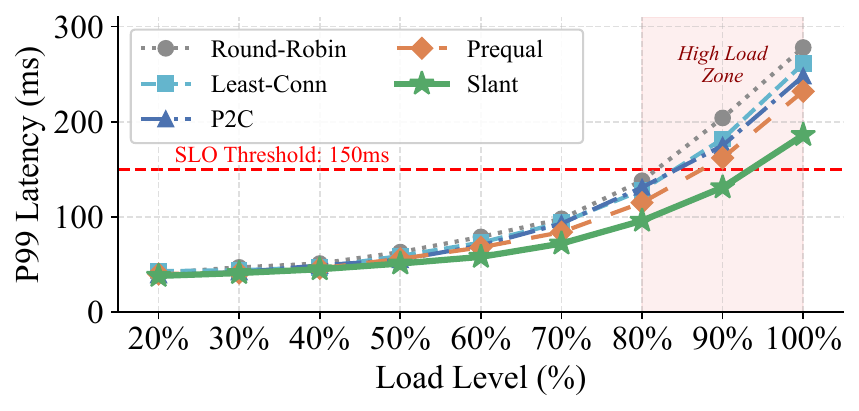} 
    \caption{P99 latency across load levels. Slant maintains lower latency than baselines, with the gap widening at high load due to better hot-cold pod management.}
    \label{fig:load-latency}
\end{figure}

Figure~\ref{fig:load-latency} shows that improved efficiency does not come at the cost of latency. Slant maintains equal or lower P99 latency across all load levels. The latency advantage is most pronounced at high load (80--100\%), where Slant's P99 latency is 15--20\% lower than baselines. This occurs because Slant keeps cold pods genuinely cold, allowing them to absorb burst traffic without the queuing delays that occur when all pods are moderately loaded.

\subsection{Robustness under Anomalies}
\label{subsec:robustness}






\subsubsection{Pod Failure Handling}

We simulate pod failures by terminating pods handling active traffic and measure recovery behavior.


\begin{figure}[t]
    \centering
    \includegraphics[width=\columnwidth]{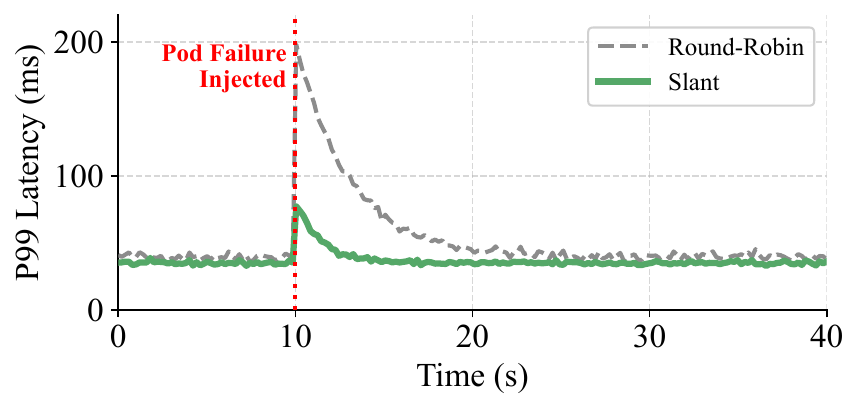} 
    \caption{Recovery from pod failure. Slant recovers 40\% faster because cold pods can immediately absorb the failed pod's traffic without becoming overloaded.}
    \label{fig:pod-failure}
\end{figure}

Figure~\ref{fig:pod-failure} shows the recovery timeline. When a hot pod fails, Slant detects the failure within 2 seconds (via health check failure and connection errors) and redistributes its traffic to other pods. Because cold pods have substantial spare capacity, they absorb the additional load without significant latency increase. Full recovery (latency returning to baseline) takes 8 seconds for Slant compared to 14 seconds for Round-Robin, a 40\% improvement.

Notably, when a cold pod fails in Slant, the impact is minimal. The system simply promotes another lightly-loaded pod to cold status and continues operation with negligible disruption.

\subsubsection{Downstream Latency Increase}

Microservices often depend on external services whose performance may degrade unexpectedly. We inject a 2$\times$ latency increase in a downstream database service and observe the cascade effects.

When downstream latency increases, requests take longer to complete, causing connection and thread pools to fill up. Traditional load balancers, unaware of this internal state, continue routing requests uniformly, potentially overwhelming all pods. Slant's multi-dimensional monitoring detects the increased queue depth and connection count, triggering proactive traffic reduction to affected pods before they become completely unresponsive.

In our experiment, Slant limits the latency increase to 1.8$\times$ (compared to 2.6$\times$ for Round-Robin) and prevents the cascading failure that causes 5\% error rate in baseline systems.

\subsection{System Overhead}
\label{subsec:overhead}

\begin{table}[t]
    \centering
    \caption{System overhead breakdown. Slant introduces minimal overhead that is negligible compared to typical request processing times.}
    \label{tab:overhead}
    \small
    \begin{tabular}{lcc}
        \toprule
        \textbf{Component} & \textbf{Metric} & \textbf{Overhead} \\
        \midrule
        \multirow{2}{*}{Control Plane} & CPU usage & 0.3 cores \\
        & Memory usage & 256 MB \\
        \midrule
        \multirow{2}{*}{Data Plane} & Per-request latency & 0.08 ms \\
        & Memory per sidecar & 12 MB \\
        \midrule
        \multirow{2}{*}{Monitoring} & Metrics bandwidth & 2.1 KB/s/pod \\
        & Weight sync bandwidth & 0.5 KB/s/pod \\
        \bottomrule
    \end{tabular}
\end{table}

Table~\ref{tab:overhead} summarizes the overhead. The control plane consumes 0.3 CPU cores and 256MB memory, which is negligible for a cluster managing thousands of pods. The data plane adds only 0.08ms per-request latency (measured as the difference between Slant and a pass-through proxy), which is less than 0.1\% of typical request latency. The additional memory per sidecar (12MB for weight tables and routing logic) is minimal compared to Envoy's base footprint. Network overhead for metrics collection and weight synchronization totals less than 3KB/s per pod, insignificant compared to application traffic.

\subsection{Ablation Study}
\label{subsec:ablation}

To understand the contribution of each component, we evaluate Slant variants with specific features disabled.

\begin{table}[t]
    \centering
    \caption{Ablation study results on Social Network at 60\% load. Each row disables one component, showing its impact on key metrics.}
    \label{tab:ablation}
    \small
    \setlength{\tabcolsep}{4pt}
    \begin{tabular}{lccc}
        \toprule
        \textbf{Config} & \textbf{P99 (ms)} & \textbf{CPU Cores} & \textbf{SLO Viol.} \\
        \midrule
        Slant (full) & 128 & 81.5 & 0.02\% \\
        \midrule
        w/o Imbalance & 142 & 110 & 0.08\% \\
        w/o Multi-dim & 135 & 82.5 & 0.15\% \\
        w/o Profiler & 138 & 96 & 0.12\% \\
        w/o Fallback & 131 & 81.5 & 0.18\% \\
        \bottomrule
    \end{tabular}
\end{table}

Table~\ref{tab:ablation} presents the ablation study results on Social Network. \textbf{Without Imbalanced Routing:} Replacing our imbalance-aware optimizer with capacity-proportional balanced routing eliminates the core contribution; allocated CPU cores increase significantly from 81.5 to 110.0, and P99 latency degrades by 11\%, confirming that intentional imbalance is essential for high resource efficiency. \textbf{Without Multi-dimensional Constraints:} Using only CPU utilization for capacity estimation increases SLO violations to 0.15\%. While average CPU core allocation remains low (82.5), the system operates unsafely during memory-intensive request bursts, leading to 8$\times$ more OOM events and 5$\times$ more timeout errors in I/O operations because the single-dimensional model fails to predict non-CPU capacity exhaustion. \textbf{Without Service Profiler:} Using fixed, conservative critical points (70\% CPU for all services) instead of learned profiles increases the required CPU cores to 96.0 and degrades P99 latency by 8\%, as fixed thresholds are too conservative for some services and too aggressive for others. \textbf{Without Fallback Controller:} Disabling anomaly detection and graduated fallback increases SLO violations to 0.18\%; under transient anomalies (e.g., brief load spikes or temporary downstream slowdowns), the lack of fallback mechanisms allows problems to escalate before the regular control loop can react.

%% file: 6-RelatedWork.tex
\section{Related Work}
\label{sec:relatedwork}
\paragraph{Microservice performance optimization and service-graph-aware systems.}
Several recent systems improve microservice performance by exploiting application structure or stronger resource control~\cite{Management1, scaling1, scaling2, scaling3, scaling4, scaling5, scaling6, scaling7, scaling8, scaling9, scaling10, scaling11, scaling12, scaling13, scaling14, lb7}. Autothrottle~\cite{autothrottle} studies SLO-oriented microservice resource management and uses a bi-level design to adjust CPU allocation based on the relationship between throttling and latency. MuCache~\cite{zhang2024mucache} improves performance through a general caching framework over microservice graphs. LuoShen~\cite{pan2024luoshen} highlights the importance of efficient traffic processing and consolidation in multi-tenant, multi-service edge clouds. These studies show that microservice performance depends on more than local request dispatch. However, they focus on resource allocation, caching, or gateway consolidation, rather than on how the \emph{load balancing policy itself} shapes the utilization patterns seen by downstream resource managers. In contrast, our work focuses on the coupling between traffic distribution and resource management: instead of pursuing balance by default, we intentionally create safe traffic imbalance so that hot replicas expose clearer scale-up signals and cold replicas become clearer scale-in targets.

\paragraph{Learning-based and performance-aware systems optimization.}
Another closely related direction is learning-based or data-driven performance optimization for complex systems~\cite{10.1145/3005745.3005750, bilal2020best}. Namyar \emph{et al.}~\cite{namyar2024end} highlight that learning-enabled systems should be analyzed end-to-end rather than at the model component alone, emphasizing the importance of system-level performance reasoning. Zeng \emph{et al.}~\cite{zeng2025learning} demonstrate in a production Alibaba Cloud CDN deployment that different operating conditions require different congestion-control choices, and that learned selection can outperform one-size-fits-all policies. These works reinforce a key lesson also reflected in our setting: under heterogeneous workloads and operating conditions, static policies are often suboptimal. Our approach shares this perspective, but targets a different control point. Rather than selecting transport protocols or analyzing learning-enabled pipelines, we learn service-specific critical points and use them to drive online, imbalance-aware routing decisions under multi-dimensional resource constraints.

\paragraph{Performance-aware control, mitigation, and coupled system dynamics.}
Recent work has also shown that control policies in large-scale systems cannot be designed independently from the performance signals they induce~\cite{DBLP:conf/nsdi/WooSHMRS18, DBLP:conf/sigcomm/ManousisSSS20, lei2025switchnic, sdn, DBLP:conf/osdi/MadhyasthaIPDAKV06, DBLP:conf/nsdi/KrishnaswamySMB23, wang2022rdc, ferreira2024impossibility, zhai2020check, sherry2015blindbox}. Namyar \emph{et al.}~\cite{namyar2025enhancing} improve failure mitigation by ranking actions using end-to-end performance information, while CoreSync~\cite{pardeshi2025coresync} explicitly studies the interaction between scheduling and overload control for latency-sensitive workloads. Beaver~\cite{yu2024beaver} addresses practical observability for distributed cloud services through efficient partial snapshots, improving the ability to reason about system behavior under dynamic conditions. These studies are complementary to ours in that they all recognize the importance of system interactions, feedback loops, and performance-aware control. Our work differs in both layer and objective. We focus on service-level request routing inside microservice deployments, and we show that the conventional goal of balanced traffic distribution may conflict with efficient vertical resource management. Building on this insight, our design combines service profiling, multi-dimensional safety constraints, imbalance-aware routing, and fallback control to improve both resource efficiency and tail latency.

%% file: 7-conclusion.tex
\section{Conclusion}
In this paper, we argued that balanced request distribution is often misaligned with efficient vertical resource management in microservice systems. To address this mismatch, we designed Slant, a resource-aware load balancing framework that profiles service-specific performance boundaries, computes conservative multi-dimensional capacity estimates, periodically applies intentional traffic imbalance through routing weights, and coordinates safely with existing VPA mechanisms through fallback control. Across three benchmark applications, a production trace-driven workload, and multiple anomaly scenarios, Slant reduces allocated CPU cores by up to 26\%, lowers P99 latency by up to 18\%, triggers VPA scale-up 45\% faster, completes scale-in 40\% faster, and improves robustness under pod failures and downstream slowdowns, without requiring changes to application code or the underlying autoscaler in production deployments and remains easy to adopt in Kubernetes clusters.